\documentclass[a4,11pt]{article}
\usepackage{latexsym,enumerate}
\usepackage{amsmath,amsthm,amsopn,amstext,amscd,amsfonts,amssymb}
\usepackage{natbib,graphicx,fullpage}
\usepackage{subfigure}
\begin{document}

\title{Comments on\\ ``Particle Markov chain Monte Carlo''\\ by
  C.~Andrieu, A.~Doucet, and R.~Hollenstein}
\author{Pierre Jacob,${}^{\dag\ddag}$ 
	Nicolas Chopin,${}^\dag$ 
	Christian P. Robert,${}^{\dag\ddag}$ 
	and H{\aa}vard Rue${}^\star$\\ 
	${}^\dag$CREST-INSEE,
	${}^\ddag$CEREMADE, Universit\'e Paris Dauphine 
	and ${}^\star$NTNU, Trondheim}

\maketitle

\begin{abstract}
This note merges three discussions written by all or some of the above authors
about the Read Paper ``Particle Markov chain Monte Carlo'' by
C.~Andrieu, A.~Doucet, and R.~Hollenstein, presented on October 16 at the Royal
Statistical Society and to be published in the {\em Journal of the Royal
Statistical Society Series B.}
\end{abstract}

\section{Introduction}
We congratulate the three authors for opening such a new vista for running MCMC
algorithms in state-space models. Being able to devise a correct
Markovian scheme based on a particle approximation of the target
distribution is a genuine ``tour de force'' that deserves an
enthusiastic recognition! This is all the more impressive when
considering that the ratio
$$
\widehat p_\theta(x_{1:T}^\star|y_{1:T}) / \widehat p_\theta(x_{1:T}(i-1)|y_{1:T})
\eqno{(11)}
$$
is not unbiased and thus invalidate the usual importance sampling
solutions, as demonstrated by
\cite{beaumont:cornuet:marin:robert:2009}. Thus, the resolution of
simulating by conditioning on the lineage truly is an awesome
resolution of the problem!

\section{Implementation and code}
We implemented the PHM algorithm for the (notoriously challenging) stochastic volatility model
$$
 y_t | x_t\sim \mathcal{N}(0, e^{x_t}),\quad x_t = \mu + \rho (x_{t-1} - \mu) + \sigma \varepsilon_t\,,
$$
based on several hundred simulated observations. With parameter moves
\begin{align*}
\mu^* &\sim \mathcal{N}(\mu, 20^{-2}), \\
\rho^* &\sim \mathcal{N}(\rho, 20^{-2}), \\
\log{\sigma^*} &\sim \mathcal{N}(\log \sigma, 20^{-2}),
\end{align*} 
and state-space moves derived from the AR(1) prior, we obtained good mixing properties
with no calibration effort, using $N=10^2$ particles and $10^4$
Metropolis--Hastings iterations, as demonstrated by Figures \ref{param} and \ref{acf}. 
{\begin{figure}
 \centering
 \includegraphics[width=0.5\textwidth]{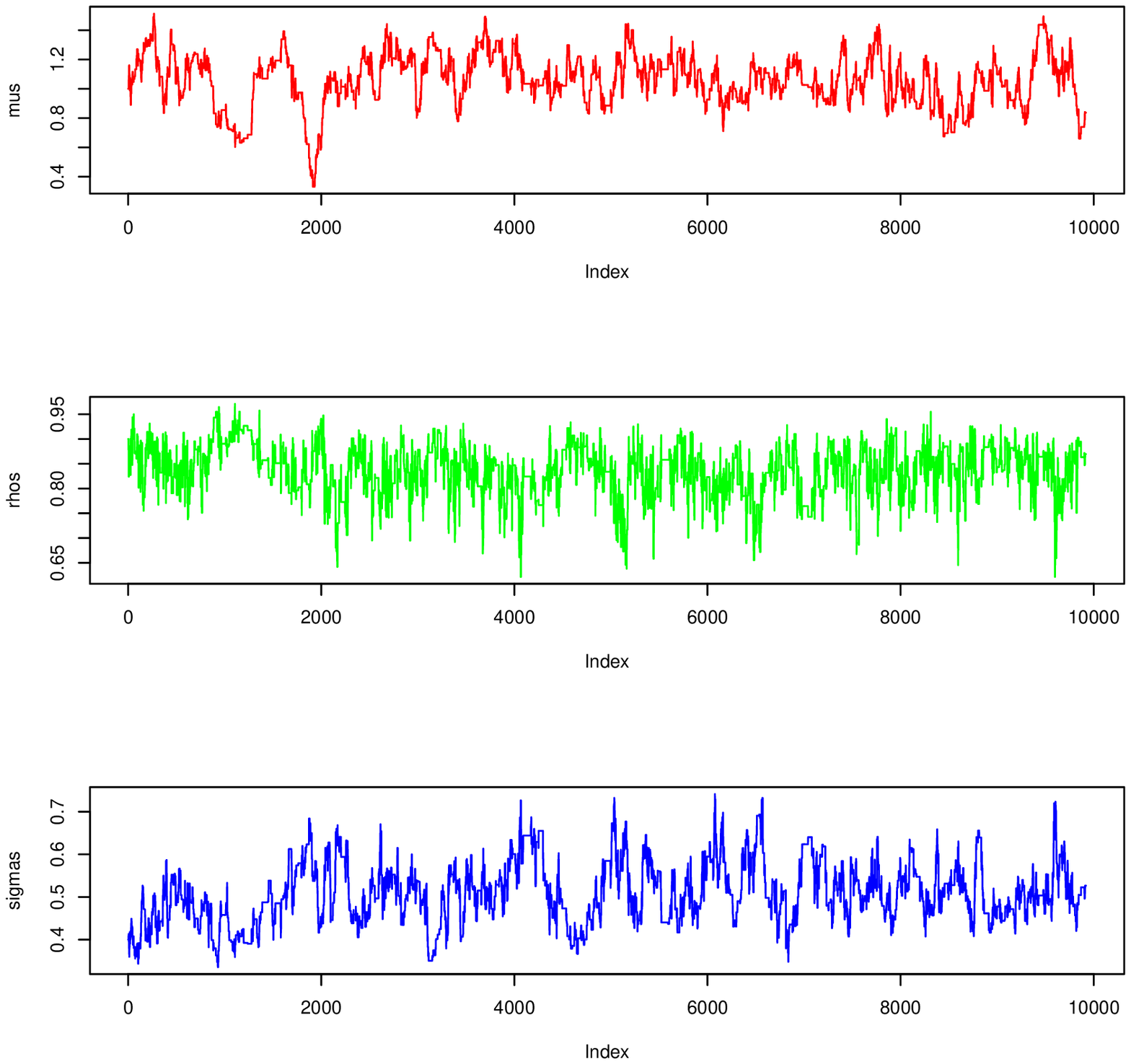}\includegraphics[width=0.45\textwidth]{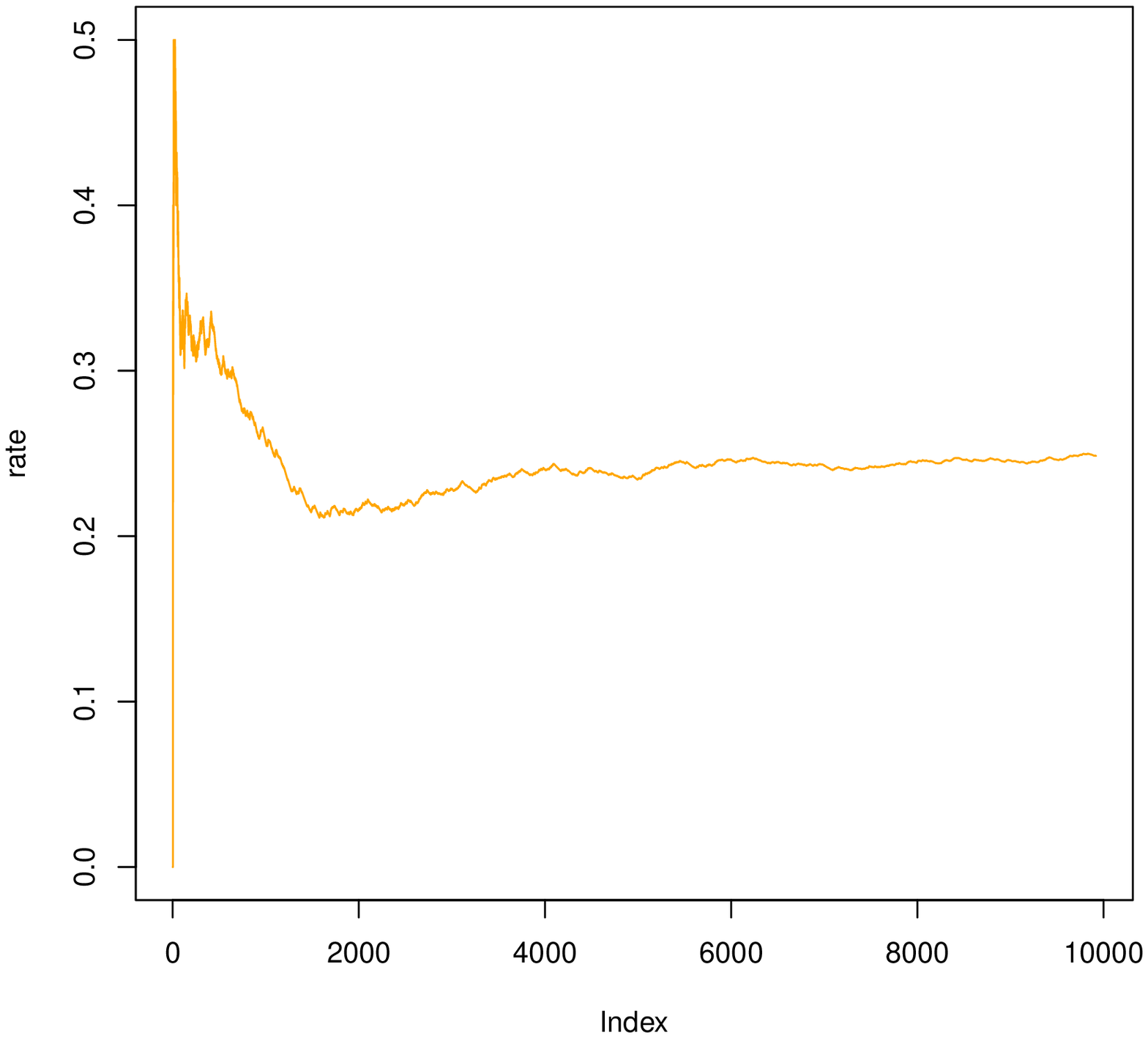}
\caption{\label{param} Evolution of the {\em (left)} parameter simulations for $\mu$, $\rho$ and $\sigma$, 
plotted against iteration indices and of the {\em (right)} estimated acceptance rate of the PMCMC
algorithm, obtained $N=10^2$ particles and $10^4$
Metropolis--Hastings iterations and a simulated sequence of $500$ observations with true values
$\mu_0=1$, $\rho_0=0.9$ and $\sigma_0=0.5$.}
\end{figure}}

{\begin{figure}
 \centering
 \includegraphics[width=0.5\textwidth]{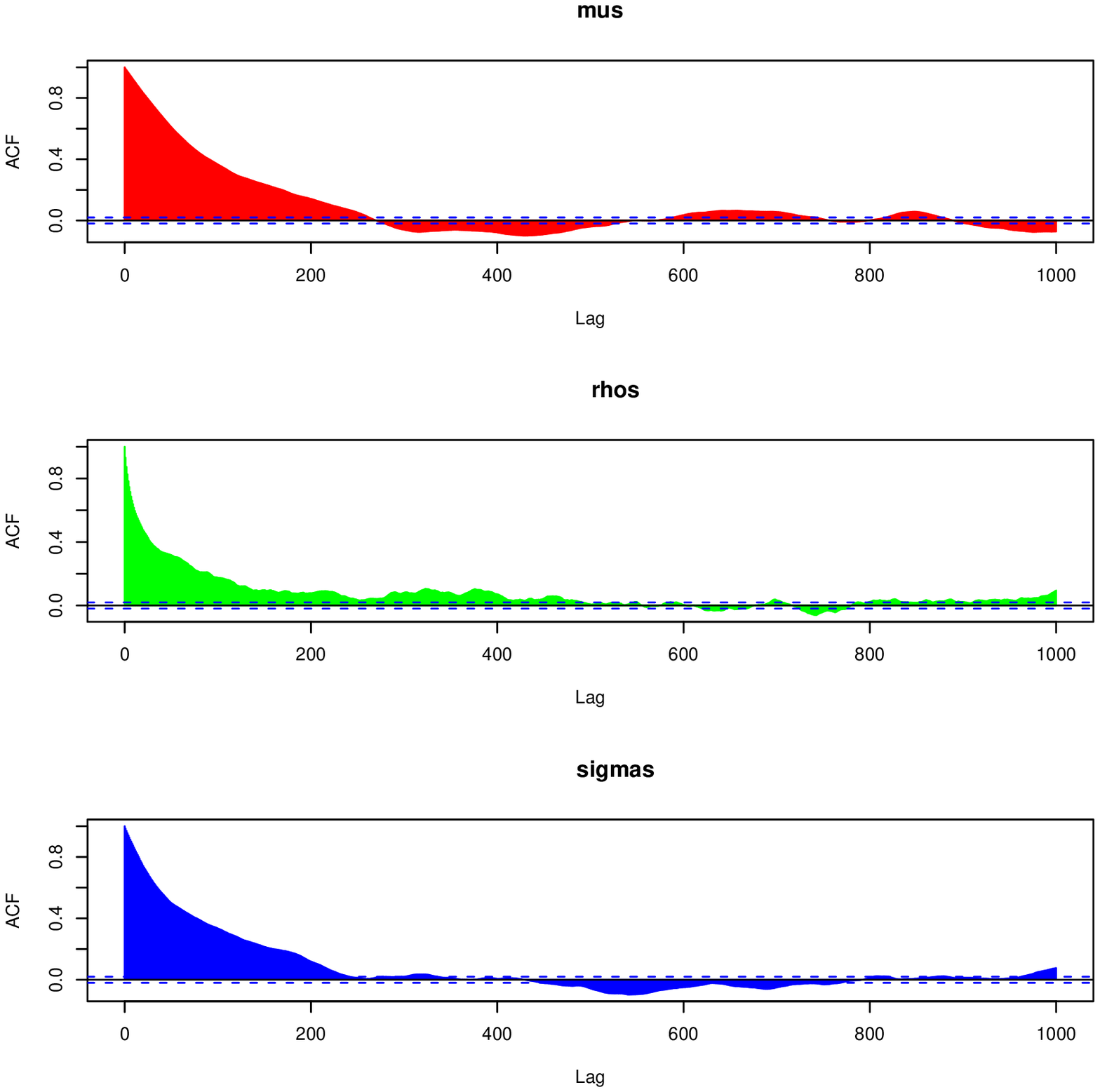}\includegraphics[width=0.5\textwidth]{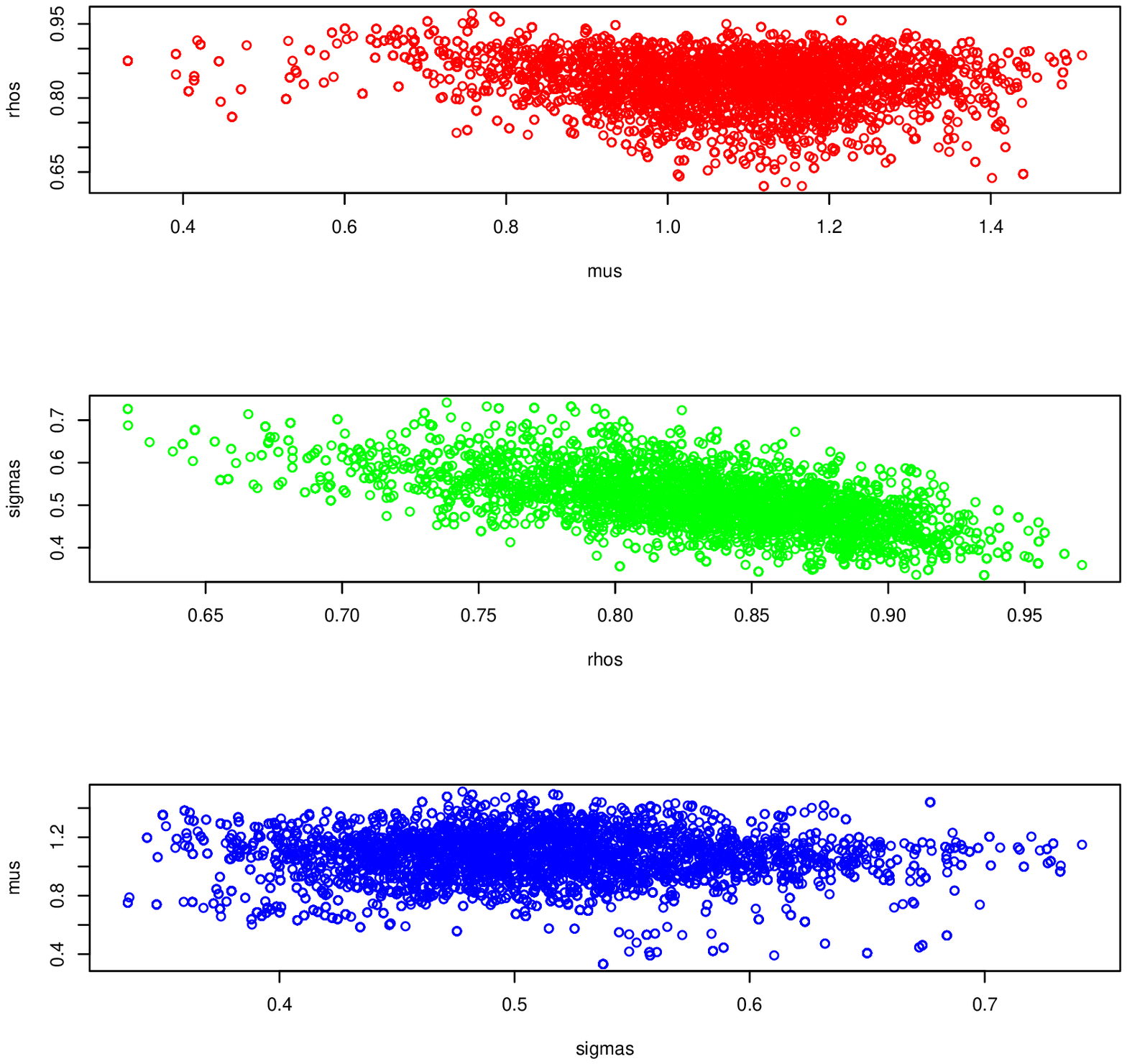}
\caption{\label{acf} Autocorrelation {\em (left)} and pairwise {\em (right)}
graphs for the $\mu$, $\rho$ and $\sigma$ sequences for the same target as in Figure \ref{param}.}
\end{figure}}

The acceptance rate for this configuration, when using a variance of $0.05^2$ for each parameter move, 
and $100$ particles for $500$ observations, was around 25\%.  With $100$ observations and $100$ particles, 
the results of the PHM algorithm showed a bimodality in the Markov chain as presented in Figure \ref{d100}.

{\begin{figure}
 \centering
\subfigure[Parameter values.]{\label{param2}\includegraphics[width=0.4\textwidth]{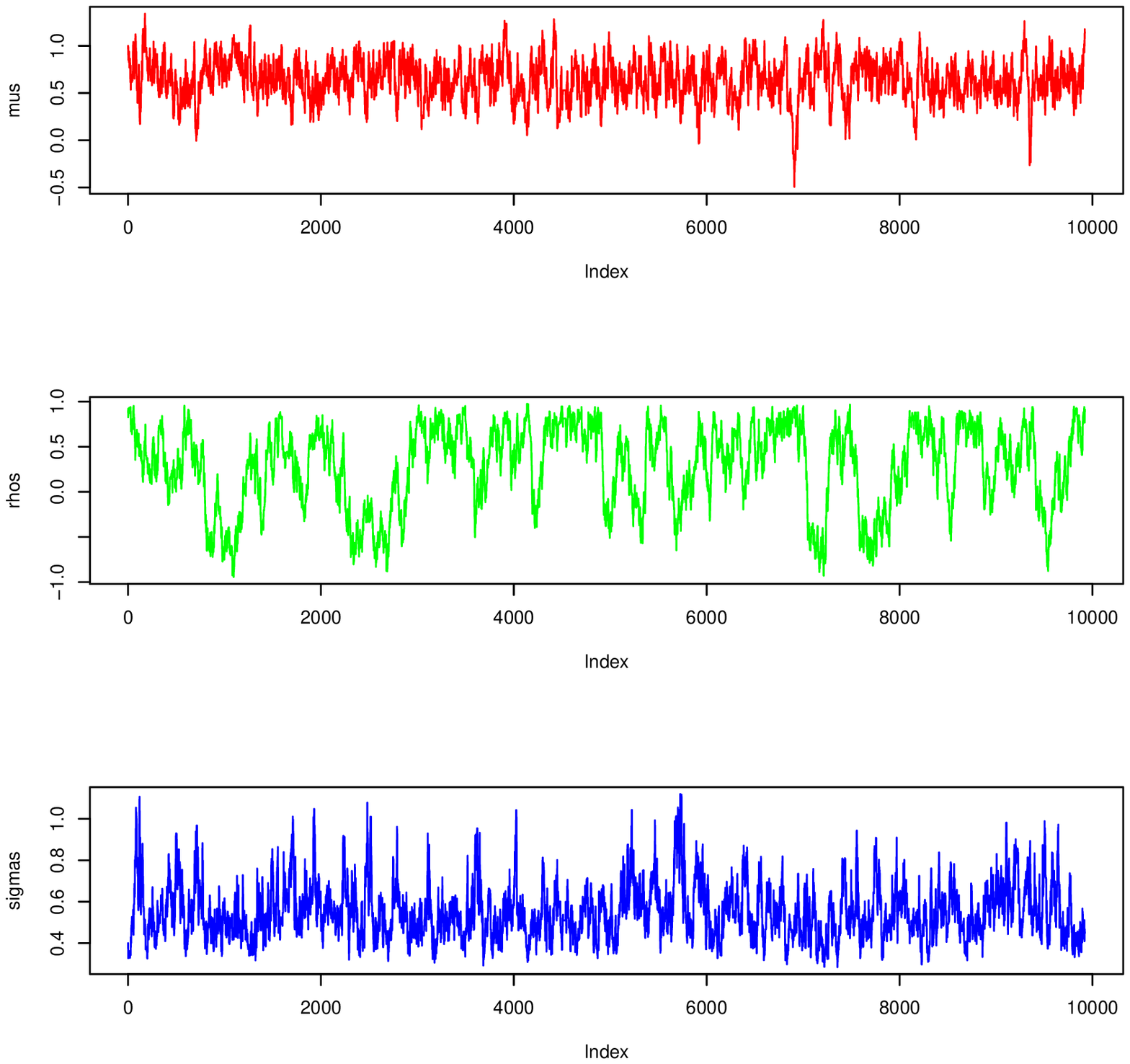}}
\subfigure[Autocorrelograms.]{\label{acf2}\includegraphics[width=0.4\textwidth]{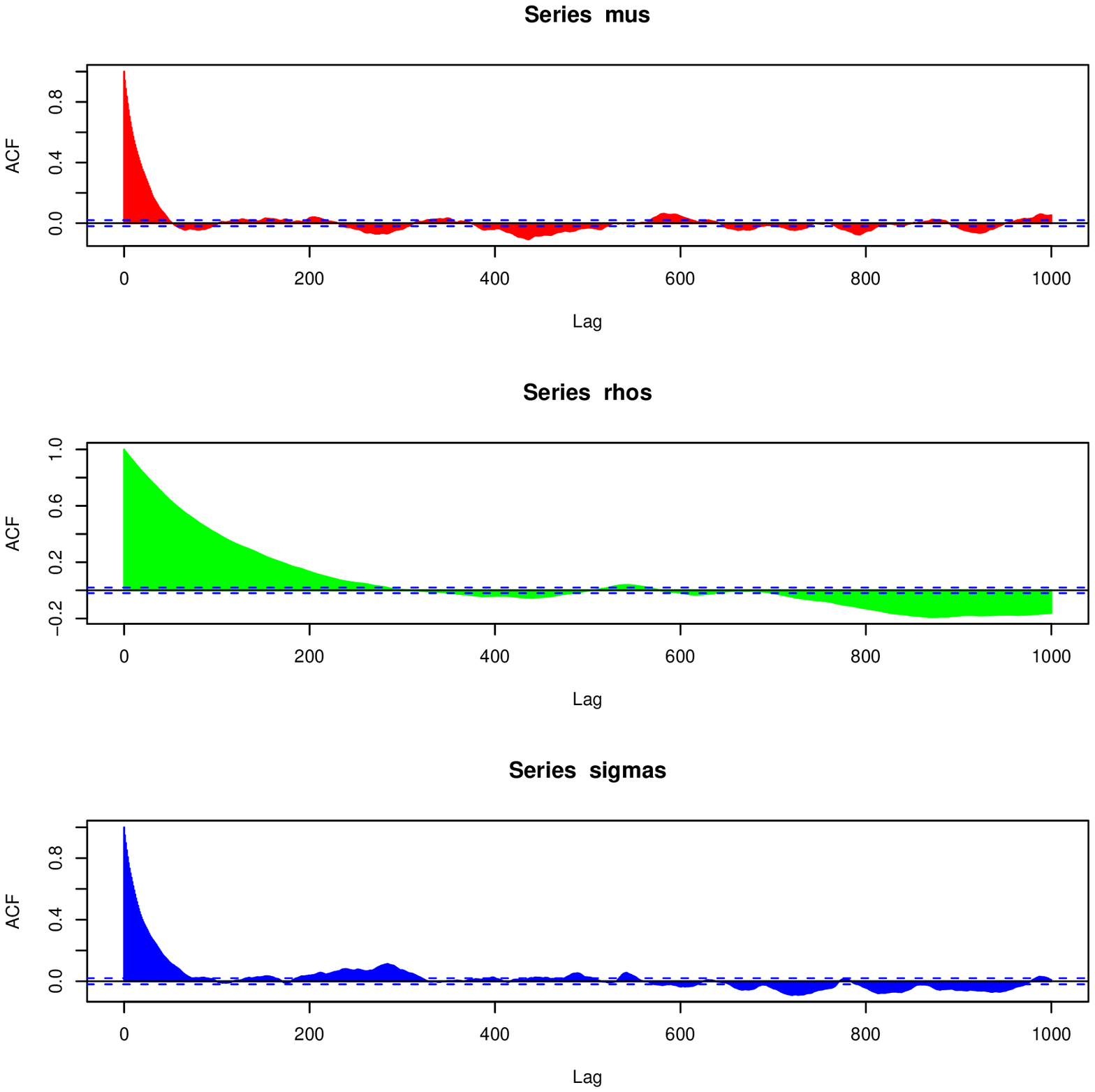}}
\caption{\label{d100} Evolution of the {\em (left)} parameter simulations for $\mu$, $\rho$ and $\sigma$, plotted against iteration indices,
and autocorrelograms for $\mu$, $\rho$ and $\sigma$, for $100$ observations and $100$ particles.}
\end{figure}}

The sequence of the $\rho$'s on the lhs of Figure \ref{d100} exhibits concentrations
around both $0.9$ and $-0.9$. This bimodality of the posterior on $\rho$ disappears when the number of 
observations grows, as shown by Figures \ref{acf} and \ref{d1000}, obtained with $1000$ 
simulated observations and $1000$ particles.

{\begin{figure}
 \centering
\subfigure[Parameter values.]{\label{param3}\includegraphics[width=0.4\textwidth]{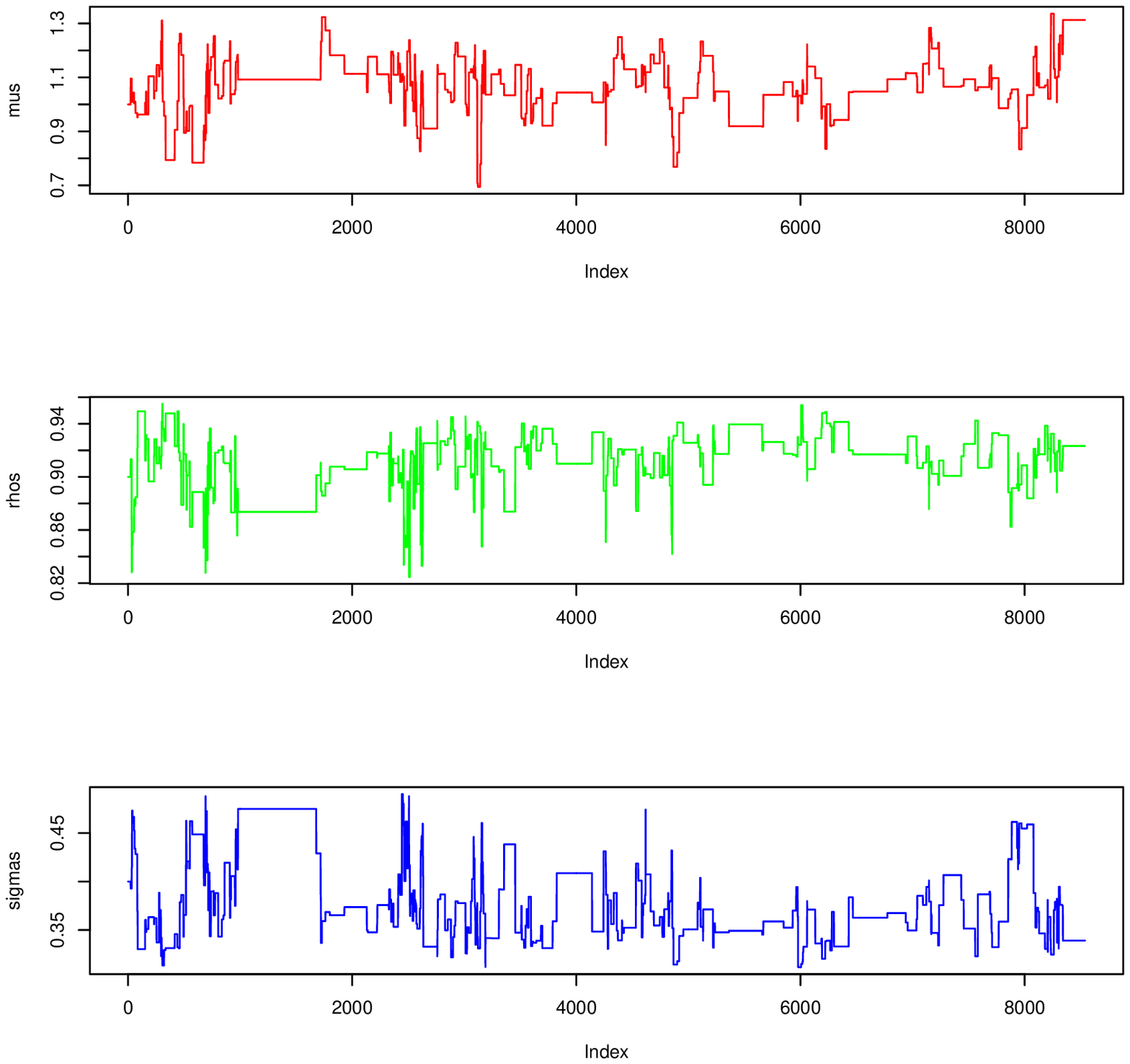}}
\subfigure[Autocorrelograms.]{\label{acf3}\includegraphics[width=0.4\textwidth]{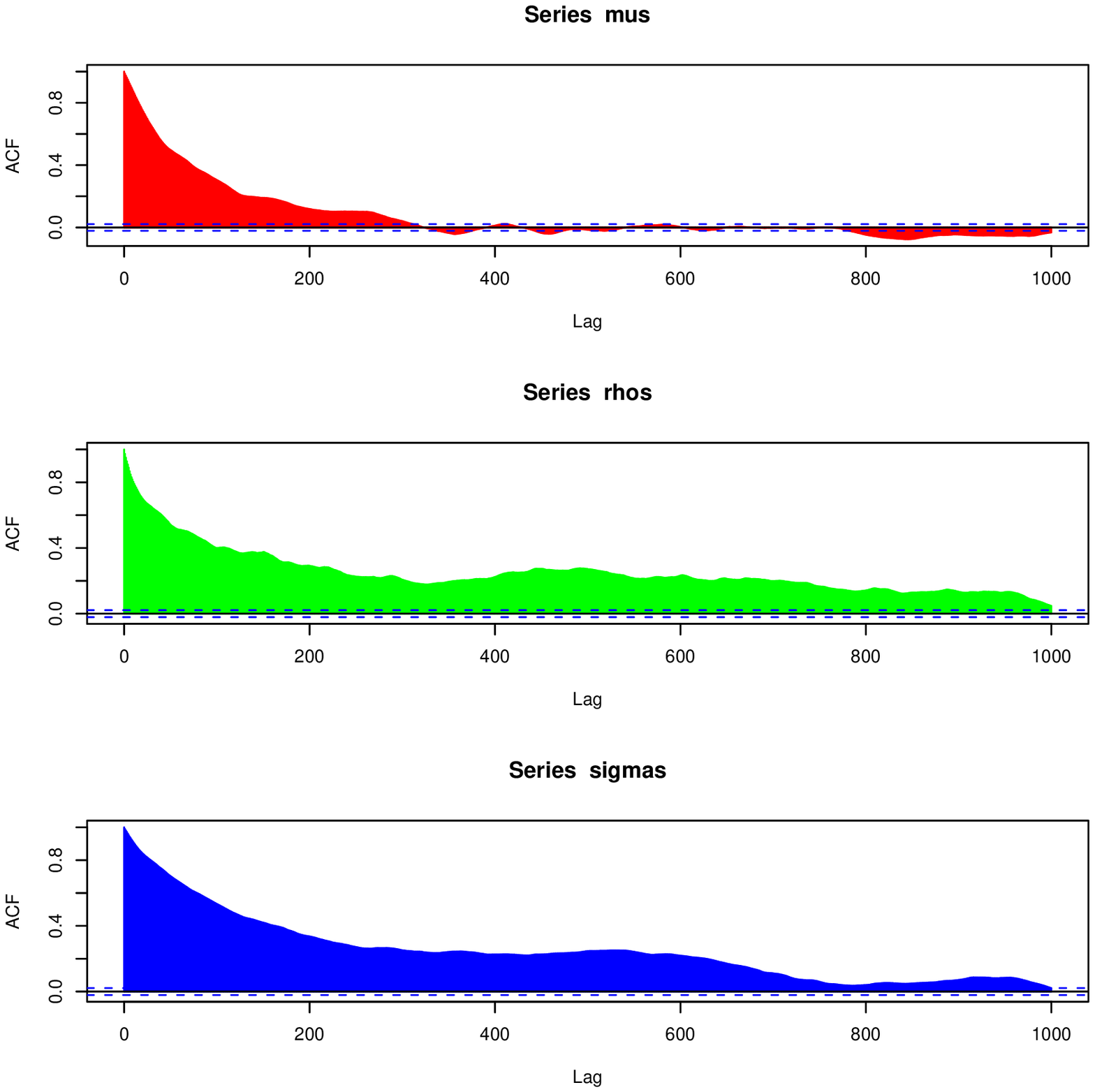}}
\caption{\label{d1000} Evolution of the {\em (left)} parameter simulations for $\mu$, $\rho$ and $\sigma$,
 and autocorrelograms for $\mu$, $\rho$ and $\sigma$, plotted against iteration indices, 
for $1000$ observations and $1000$ particles.
}
\end{figure}}

Bimodality of the posterior on $\rho$ could possibly occur for a larger number of observations. It may also be an 
artifact of the simulation method, which would require a higher number of particles or of iterations to assess. 
Unfortunately, this is computationally very demanding: using our Python programme on
$10,000$ iterations, $1000$ observations and $1000$ particles requires five hours on a mainframe computer.

Figures \ref{d100} and \ref{d1000}, obtained with the same proposal variance, also illustrate the severe decrease in the 
acceptance rate when the number of observations grows: the proposed parameter values get rejected for more than 100 iterations
in a row on Figure \ref{d1000}.

Our computer program is written in the Python language. It is available for download at\hfil 
\verb+http://code.google.com/p/py-pmmh/+
and may be adapted to any state-space model by simply rewriting two lines of the code, namely those 
that (a) compute $p(y_{t}|x_t)$ and (b) simulate $x_{t+1}|x_t$.
Contemplating a different model does not even require the calculation of full conditionals, in contrast to Gibbs sampling. 
Another advantage of the \verb+py-pmmh+ algorithm is that it is trivial to parallelise. (Adding a comment before the loop over 
the particle index is enough, using the \verb+OpenMP+ technology.)

\section{Recycling}
Although this is already addressed in the last section of the paper,
we mention here possible options for a better recycling of the numerous simulations produced by the algorithm.
This dimension of the algorithm deserves deeper study, maybe to the extent of allowing for a finite time horizon
overcoming the MCMC nature of the algorithm, as in the PMC solution
of \cite{cappe:douc:guillin:marin:robert:2007}. 

A more straightforward remark is that, due to the additional noise brought by the resampling
mechanism, more stable recycling would be produced both in the
individual weights $\omega_n(X_{1:n})$ by Rao--Blackwellisation of the
denominator in eqn.~(7) as in \cite{iacobucci:marin:robert:2008} and
over past iterations by a complete reweighting scheme like AMIS
\citep{cornuet:marin:mira:robert:2009}. Another obvious question is
whether or not the exploitation of the wealth of information provided
by the population simulations is manageable via adaptive MCMC methods
\citep{andrieu:robert:2001,roberts:rosenthal:2006}

\section{Model choice}
Since
$$
\widehat p_\theta(y_{1:T}) = \widehat p_\theta(y_1) \prod_{n=2}^T \widehat p_\theta(y_n|y_{1:n-1}) 
$$
is an unbiased estimator of $p_\theta(y_{1:T})$, there must be direct
implications of the method towards deriving better model choice
strategies in state-space models, as exemplified in Population Monte Carlo by
\cite{kilbinger:wraith:robert:benabed} in a cosmology setting.

In fact, the paper does not cover the corresponding calculation of the
marginal likelihood $p(y)$, the central quantity in model choice. 
However, the PMCMC approach seems to lend itself naturally to the use
of  Chib's (\citeyear{chib:1995}) estimate, i.e. 
$$ p(y) = \frac{p(\theta^\star)p(y|\theta^\star)}{p(\theta^\star| y)} $$
for any $\theta^\star$. Provided the $p(\theta| x,y )$ density 
allows for a closed-form expression, the denominator may be estimated by 
\begin{equation*}
p(\theta| y) = \int p(\theta| x,y) p(x| y)\,\text{d}x \approx \frac{1}{M}\sum_{i=1}^M p(\theta| 
x=x_{i},y)\,,
\end{equation*}
where the $x_i$'s, $i=1,...,M,$ are provided by the MCMC output. 

The novelty in Andrieu et al. (2009) is that $p(y|\theta)$ in the numerator needs to be
evaluated as well. Fortunately, as pointed out above, each iteration $i$ of the PMCMC sampler
provides a Monte Carlo estimate of $p(y|\theta=\theta_i)$, where $\theta_i$ is the current parameter
value. Some care may be required when choosing $\theta^\star$; e.g. selecting the $\theta^\star$ with largest
(evaluated) likelihood may lead to a biased estimator.

We did some experiments in order to compare the approach described
above with INLA \citep{rue:martino:chopin:2009} and nested sampling 
(\citealp{skilling:2007b}; see also \citealp{chopin:robert:2007}), using the stochastic
volatility example of Rue et al. (2009). Unfortunately, our PMCMC
program requires more than one day to complete (for a number $N$ of
particles and a number $M$ of iterations that are sufficient for
reasonable performance), so we were unable to include the results in
this discussion. A likely explanation is that the cost of PMCMC is at
least $O(T^2)$, where $T$ is the sample size ($T=945$ in this
example), since, according to the authors, good performance requires
than $N=O(T)$, but our implementation may be sub-optimal as well.

Interestingly, nested sampling performs reasonably well on this example
(reasonable error obtained in one hour), and, as reported by \cite{rue:martino:chopin:2009},
the INLA approximation is  fast (one second)
and very accurate, but more work is required for a more meaningful 
comparison. 

\section{Connections with some of the earlier literature}
Two interesting metrics for the impact of a Read Paper are: (a)
the number of previous papers it impacts in some way;
and (b) the number of interesting theoretical questions it opens.
In both respects, this paper fares very well. 

Regarding (a), in many complicated models the only tractable operations
are state filtering and likelihood evaluation, as shown, e.g., in the
continuous-time model of \cite{chopin:varini:2007}. In such settings,
the PHM algorithm offers Bayesian estimates ``for free'', which is very nice. 

Similarly, \cite{chopin:2007}, see also \cite{fearnhead:liu:2007}, formulates
change-point models as state-space models, where the state
$x_{t}=(\theta_t,d_t)$, comprises the current parameter $\theta_t$
and the time since the last change  $d_{t}$. In this setting, one may 
use SMC to recover the trajectory $x_{1:T}$, i.e., all the change
dates and parameter values. It works well when $x_t$ forgets about its past
quickly enough, but this forbids hierarchical priors for the
durations and the parameters.  PHM removes this
limitation: Chopin's (2007) SMC algorithm may indeed be embedded 
into a PHM algorithm, where each iteration corresponds to a different
value of the hyper-parameters.  This comes as a cost however, as each MCMC
iteration need run a complete SMC algorithm.

Regarding point (b), several questions, which have already been answered in
the standard SMC case, may again be asked about PMCMC:  Does residual
resampling outperform multinomial resampling? Is the algorithm with
$N+1$ particles strictly better than the one with $N$ particles? What
about Rao-Blackwellisation, or the choice of the proposal
distribution?  One technical difficulty is that marginalising out
components always reduces the variance in SMC, but not in MCMC. Another
difficulty is that PMCMC retains only one particle trajectory
$x_{1:T}$, hence the impact in reducing variability between particles
is less immediate. 

Similarly, obtaining a single trajectory $x_{1:T}$ from a forward filter
is certainly much easier than obtaining many of them, but this may still be
quite demanding in some scenarios, i.e., there may be so much degeneracy
in $x_1$ that not a single particle does contain a $x_1$ that is in the support of
$p(x_{1}|y_{1:T})$.

\section*{Acknowledgements}
Pierre Jacob is supported by a PhD Fellowship from the AXA Research Fund.
Nicolas Chopin and Christian Robert are partly supported by the Agence Nationale de la Recherche
(ANR, 212, rue de Bercy 75012 Paris) through the 2009-2012 projects {\sf Big'MC}.


\end{document}